\documentstyle[12pt,epsf]{article}
\def\singlespace {\smallskipamount=3.75pt plus1pt minus1pt
                  \medskipamount=7.5pt plus2pt minus2pt
                  \bigskipamount=15pt plus4pt minus4pt
                  \normalbaselineskip=15pt plus0pt minus0pt
                  \normallineskip=1pt
                  \normallineskiplimit=0pt
                  \jot=3.75pt
                  {\def\smallskip {\vskip\smallskipamount}}
                  {\def\medskip   {\vskip\medskipamount}}
                  {\def\bigskip   {\vskip\bigskipamount}}
                  {\setbox\strutbox=\hbox{\vrule
                    height10.5pt depth4.5pt width 0pt}}
                  \parskip 7.5pt
                  \normalbaselines}
\def\middlespace {\smallskipamount=5.825pt plus1.5pt minus1.5pt
                  \medskipamount=11.25pt plus3pt minus3pt
                  \bigskipamount=22.5pt plus6pt minus6pt
                  \normalbaselineskip=22.5pt plus0pt minus0pt
                  \normallineskip=1pt
                  \normallineskiplimit=0pt
                  \jot=5.825pt
                  {\def\smallskip {\vskip\smallskipamount}}
                  {\def\medskip   {\vskip\medskipamount}}
                  {\def\bigskip   {\vskip\bigskipamount}}
                  {\setbox\strutbox=\hbox{\vrule
                    height15.75pt depth6.75pt width 0pt}}
                  \parskip 7.25pt
                  \normalbaselines}
\def\dblspc {\smallskipamount=7.5pt plus2pt minus2pt
                  \medskipamount=15pt plus4pt minus4pt
                  \bigskipamount=30pt plus8pt minus8pt
                  \normalbaselineskip=30pt plus0pt minus0pt
                  \normallineskip=2pt
                  \normallineskiplimit=0pt
                  \jot=7.5pt
                  {\def\smallskip {\vskip\smallskipamount}}
                  {\def\medskip   {\vskip\medskipamount}}
                  {\def\bigskip   {\vskip\bigskipamount}}
                  {\setbox\strutbox=\hbox{\vrule
                    height21.0pt depth9.0pt width 0pt}}
                  \parskip 15.0pt
                  \normalbaselines}

\def\dg{\dagger }
\def\pr{\prime }

\def\be{\begin{equation}}
\def\lan{\left\langle}
\def\ran{\right\rangle}

\def\j-{\J_-}

\def\ee{\end{equation}}

\def\bearr{\begin{eqnarray}}
\def\bearrs{\begin{eqnarray*}}
\def\eearr{\end{eqnarray}}
\def\eearrs{\end{eqnarray*}}
\def\barr{\begin{array}}
\def\earr{\end{array}}

\def\nn8{\nonumber\\[10pt]}
\def\l{\left}
\def\r{\right}

\def\dis{\displaystyle}

\def\ed{\end{document}}

\def\cod{{\cal O}^\dagger}
\def\co{{\cal O}}
\def\ce{{\cal E}}
\def\cf{{\cal F}}
\begin{document}
\singlespace
\begin{center}
{\bf Nature of matrix elements in the quantum chaotic
domain of \\ interacting particle systems} 
\vskip 0.25cm

V.K.B. Kota$^{a,}$ \footnote{Invited talk in the  National Conference on {\it
`Dynamical Systems: Recent Developments'} held at 
Unversity of Hyderabad, Hyderabad during 4-6th November, 1999.} 
and R. Sahu$^{b}$ \\
$^{a}${\it Physical Research Laboratory, Ahmedabad \,\,380 009, India} \\
$^{b}${\it Physics Department, Berhampur University,
Berhampur\,\, 760 007, India}

\end{center}
\vskip 0.4cm

{\small
\noindent {\bf ABSTRACT:} There is a newly emerging
understanding that in the chaotic domain of isolated finite
interacting many particle systems smoothed densities define the
statistical description of these systems and these densities
follow from embedded (two-body) random matrix ensembles and
their various deformations. These ensembles predict that the
smoothed form of matrix elements of a transition operator
between the chaotic eigenstates weighted by the densities at the
two ends (i.e. the bivariate strength density) will be a
bivariate Gaussian with the bivariate correlation coefficient
arising out of the non-commutability of the hamiltonian and the
transition operator involved. The ensemble theory extends to
systems with a mean-field and a chaos generating two-body
interaction (as in nuclei, atoms and diffusive quantum dots).
These developments in many-body quantum chaos are described with
special reference to one-body transition operators.

\begin{center}
{\bf 1. Introduction }
\end{center}

\noindent Random matrix physics, started with Wigner and Dyson's GOE, GUE and
GSE hamiltonian random matrix ensembles \cite{Br-81}, has grown far and wide in
the last fifteen years with the investigation of a variety of random matrix
ensembles (interpolating, banded, partitioned, chiral etc.) and applications in
almost all branches of physics \cite{Gu-98} (most recent application being in
Econophysics \cite{St-99}). As Guhr et al \cite{Gu-98} state `the universal
validity of random matrix theories is reminiscent of the  success of
thermodynamics in the last century $\cdots$'.  Recently it is recognized by 
large number of research groups in atomic, molecular, nuclear and mesoscopic
physics (see for example \cite{Fl-99,Ks-98,Ko-99,Al-97}), in the context
of many-body quantum chaos, that embedded random matrix ensembles (EE) (in
particular EGOE($k$), the embedded Gaussian orthogonal ensemble of random
matrices of $k$-body interactions) are relavant for `isolated finite
interacting particle systems'. Our purpose in this article is to discuss sum of
the results of EE for transition matrix elements.  In Section 2 definition and
construction of EGOE(2) for fermion systems are given; introduced here are also
various deformed EGOE. The EGOE(k) results for density of states
and transition matrix elements are briefly described in Section 3.  The topic
`quantum chaos and transition strength sums' form Section 4. In Section 5,
EE(1+2) results for matrix elements of one-body transition operators with
emphasis on the role of the mean-field basis and strength functions are
presented. Section 6  gives concluding remarks.

\begin{center}
{\bf 2. Embedded Ensembles}
\end{center}

\noindent EGOE($k$) for many fermion systems is defined by assuming
that the many particle space is a direct product space, of
single particle states, as in the nuclear/atomic shell model. 
EGOE($k$) for $m$ ($m > k$) fermion
systems (with the particles distributed say in $N$ single
particle states) is generated by defining the hamiltonian $H$,
which is say $k$-body, to be GOE in $k$-particle space and then
propagating it to $m$ - particle spaces by using the geometry of
the $m$ - particle spaces \cite{Br-81}. Let us consider EGOE(2). Given the
single particle states $\l| \l. \nu_i \ran \r.$,
$i=1,2,\ldots,N$, operator form of the 2-body hamiltonian is (with $a$ in $\lan
\;|\; |\;\ran_a$ indicating that the states involved are antisymmetrized),
\be
H = \dis\sum_{\nu_i < \nu_j,\;\nu_k < \nu_l} \lan \nu_k\;\nu_l
\mid H \mid \nu_i\;\nu_j\ran_a\;a^\dg_{\nu_l}\,a^\dg_{\nu_k}\,
a_{\nu_i}\,a_{\nu_j}
\ee
Symmetries for the two-body matrix elements 
(TBME) $\lan \nu_k\;\nu_l \mid H \mid \nu_i\;\nu_j\ran_a$ in (1) are,
\be
\barr{c}
\lan \nu_k\;\nu_l \mid H \mid \nu_j\;\nu_i\ran_a = - \lan \nu_k\;\nu_l
\mid H \mid \nu_i\;\nu_j\ran_a \;\;,\;\;\;\;\;
\lan \nu_k\;\nu_l \mid H \mid \nu_i\;\nu_j\ran_a = \lan \nu_i\;\nu_j
\mid H \mid \nu_k\;\nu_l\ran_a  
\earr
\ee
The hamiltonian $H$ in $m$-particle spaces is defined in
terms of the TBME via the direct product structure. 
The non-zero matrix elements of the $m$-particle $H$ matrix are of three types,
\be
\barr{l}
\lan \nu_1 \nu_2 \cdots \nu_m \mid H \mid \nu_1 \nu_2 \cdots
\nu_m \ran_a = \dis\sum_{\nu_i < \nu_j < \nu_m}\; \lan \nu_i
\nu_j \mid H \mid \nu_i \nu_j \ran_a \nn8
\lan \nu_p \nu_2 \nu_3 \cdots \nu_m \mid H \mid \nu_1 \nu_2 \cdots
\nu_m \ran_a = \dis\sum_{\nu_i= \nu_2}^{\nu_m}\; \lan \nu_p
\nu_i \mid H \mid \nu_1 \nu_i \ran_a \nn8
\lan \nu_p \nu_q \nu_3 \cdots \nu_m \mid H \mid \nu_1 \nu_2
\nu_3 \cdots \nu_m \ran_a = \lan \nu_p
\nu_q \mid H \mid \nu_1 \nu_2 \ran_a 
\earr
\ee
EGOE(2) is defined by (1)-(3) with GOE representation for $H$ in two-particle 
space,
\be
\barr{c}
\lan \nu_k\;\nu_l \mid H \mid \nu_i\;\nu_j\ran_a\;\; \mbox{are independent
Gaussian random variables}\nn8
\overline{\lan \nu_k\;\nu_l \mid H \mid \nu_i\;\nu_j\ran_a} = 0,\;\;\;
\overline{\l|\lan \nu_k\;\nu_l \mid H \mid \nu_i\;\nu_j\ran_a\r|^2} =
v^2(1+\delta_{(ij),(kl)})
\earr
\ee
In (4) bar denotes ensemble average and $v$ is a constant. 
Note that the $H(m)$ matrix dimension $d$ is $d(N,m)={\scriptsize{\l(\barr{c}
N \\ m\earr\r)}}$ and the number of independent matrix elements ($ime$)
are $ime(N)=d_2(d_2+1)/2$ where the two-particle space dimension $d_2$ =
$N(N-1)/2$. For example, $d(12,6)=924$ and $ime(12)=2211$.
The EGOE(2) is also called TBRE (two-body random matrix
ensemble).  Extension of (1-4) for boson systems is
straightforward and it is described elsewhere \cite{Ki-99}.
Hamiltonians for many interacting particle systems contain a
mean-field part (one-body part $h$) and a two-body residual
interaction $V$ mixing the configurations built out of the
distribution of particles in the mean-field single particle
orbits; $h$ is defined by single particle energies (SPE)
$\epsilon_i,\;i=1-N$ and $V$ is defined by TBME.  Then it is
more realistic to use EE(1+2), the embedded ensemble of
(1+2)-body hamiltonians,
\be
\mbox{EE(1+2) :}\;\;\{H\} = [h(1)] + \lambda\;\{V(2)\} \;\;.
\ee
Here $\{V\}$ is EE(2), i.e.  it is EGOE(2) with $v=1$ in (4) or
an ensemble with TBME being independent random variables with a
distribution different from Gauassian (for example uniform
distribution). Similarly $[h]$ is a fixed hamiltonian or an
ensemble with SPE chosen random but following some distribution.
Finally, $[h]$ and $\{V\}$ are independent.  It is to be
expected that the generic features of EE(1+2) approach those of
EGOE($k$) for sufficiently large values of $\lambda$ and
significant results emerge as $\lambda$ is varied starting from
$\lambda=0$.  Let us mention that EE(1+2) is also called TBRIM
(two-body random interaction model).  A second class of EE are
the partitioned embedded ensembles (p-EE) where the hamiltonian
in two-particle space is block structured (i.e. the space
divides into subspaces $\sum\;\oplus \Gamma$) with variances of the
matrix elements in each block being $v^2_{\Gamma \Gamma^\prime}$
for the block connecting $\Gamma$ and $\Gamma^\prime$ subspaces.
Third class of EE are EE-sym where $v^2_{\Gamma
\Gamma^\prime}=0$ for $\Gamma \neq \Gamma^\prime$ and these
ensembles are important for Hamiltonians carrying symmetries
(for example $J$ or $JT$ in nuclei and $J$ or $LST$ in atoms).
Finally there are also the modified $K$+EE ensembles where $K$
is a fixed operator.

\begin{center}
{\bf 3. Basic results for EGOE(k)}
\end{center}

Generic results for state densities and transition matrix
elements, for EGOE(k), that are essentially valid in the dilute
limit, which correspond to $N, m, k \rightarrow \infty$, $m/N
\rightarrow 0$ and $k/m \rightarrow 0$, are well known.
The eigenvalue density $I(E)$ or its normalized version
$\rho(E)$ takes Gaussian form \cite{Mf-75},
\be
\barr{c}
I(E) =  \langle\langle \delta (H-E) \rangle\rangle
= d\; \rho(E)\;;\;\;\;\;
\rho(E)  \stackrel{EGOE}\longrightarrow {\overline{\rho(E)}} =
\rho_{\cal G}(E) = \dis\frac{1}{\dis\sqrt{2\pi} \sigma} exp -
\dis\frac{1}{2} \l(\dis\frac{E-\epsilon}{\sigma}\r)^2
\earr
\ee
The binary correlation approximation, originally used by Wigner
for deriving the semi-circle state density for GOE ($k=m$ in EGOE(k))
is used by Mon and French \cite{Mf-75} to derive (6) via
the $m$-particle space moments $\lan H^p \ran^m$ of $I(E)$.
Firstly it is seen that by definition all odd moments of $I(E)$
will vanish. Using the normalization that $\lan H^2
\ran^{m=2}=1$ (then {\scriptsize{$\lan H^2 \ran^{m}=\l(\barr{c} m \\ k \earr
\r)$}})  one has the basic result {\scriptsize{$\overline{H(k)
O(t) H(k)} \Rightarrow \l( \barr{c} m-t \\ k \earr \r)\,O(t)$}} 
under ensemble average in $m$-particle spaces and in the
dilute limit. Using this and that in the trace $\lan H^p \ran^m$
binary associations dominate, one can drive formulas for the
moments by writing down all possible binary association
diagrams. For example for $\lan H^4 \ran$ there are three
diagrams, $
AABB \oplus ABBA \oplus ABAB \Rightarrow 2\{AABB\} \oplus ABAB$. 
Note that $A$ and $B$ are $H$-operators and they are $k$-body in
nature. Evaluating the irreduciable diagrams $AABB$ and $ABBA$
give the 4th reduced moment $\mu_4$ and the 4th cumulant $k_4$,
\be
k_4 = \mu_4 -3 = \l\{\lan H^2 \ran^m\r\}^{-2}\;\lan H^4 \ran^m \;\;-3 =
\l( \barr{c} m-k \\ k \earr \r) \l( \barr{c} m \\ k
\earr \r)^{-1} -1 \;\;\stackrel{k << m}{\longrightarrow}\;\; -k^2/m
\ee
Similar results for the 6th and 8th cumulants are derived: $
k_6=k^3(6k-1)/m^2 + O(1/m^3)$ and $ k_8=-4k^5(23k-9)/m^3 +
O(1/m^4)$. Thus in the dilute limit one recovers the Gaussian
form for state densities (note that we need in fact not $k/m
\rightarrow 0$ but $k^2/m \rightarrow 0$).  Thus, for a two-body
interaction $m \sim 12$ gives a good Gaussian. Note that for
$m=4$ one has $k_4=-1$ implying semi-circle shape as seen in
many numerical calculations. In practice one has to apply
Edgeworth (mostly 3rd and 4th moment/cumulant) corrections to
the Gaussian form. More important point, though not discussed
here, is that local level fluctuations given by EGOE(k) are of
GOE type.

Similar to (6), for EGOE(k) the transition matrix elements
weighted by the densities at the two ends (i.e. bivariate strength 
densities) take bivariate Gaussian form \cite{Fr-88},
$$
\barr{c}
I_{biv;{\cal O}}(E,E^\pr) = 
\lan\lan {\cal O}^{\dag} \delta (H - E^\pr ) {\cal O}
\delta (H-E) \ran\ran \nn8
=  I^\pr (E^\pr ) \l| \lan
E^\pr \mid {\cal O} \mid E \ran \r|^2 I(E) 
 =  \lan\lan {\cal O}^{\dag} {\cal O} \ran\ran\;
\rho_{biv;{\cal O}}(E,E^\pr)\;;
\earr
$$
\be
\barr{c}
\rho_{biv;{\cal O}}(E,E^\pr) \stackrel{EGOE}\longrightarrow 
{\overline{\rho_{biv;{\cal O}}(E,E^\pr)}} = \rho_{biv-{\cal
G}; {\cal O}}(E,E^\pr\;;\; \epsilon_1, \epsilon_2, \sigma_1,
\sigma_2, \zeta) = \nn8
\dis\frac {1} {2\pi \sigma_1\sigma_2
\sqrt {1 - \zeta^2}}\,
exp\l\{ - \dis\frac{1}{2(1 - \zeta^2)}  {\l[\l(\dis\frac {E
- \epsilon_1} {\sigma_1}\r)^2 
-2\zeta\l(\dis\frac {E - \epsilon_1} {\sigma_1}\r)
\l(\dis\frac {E^\pr - \epsilon_2} {\sigma_2}\r) + \l(\dis\frac {E^\pr
- \epsilon_2} {\sigma_2}\r)^2 \r]} \r\}
\earr
\ee
The bivariate reduced central moments of $\rho_{biv;{\cal O}}$
are $\mu_{pq} = \l. \lan {\cal O}^{\dag} \l({H - \epsilon_2
\over \sigma_2} \r)^q {\cal O} \l({H - \epsilon_1 \over
\sigma_1} \r)^p \ran \r/ \lan {\cal O}^{\dag} {\cal O} \ran$ and
$\zeta = \mu_{11}$ is the bivariate correlation coefficient.
Let us consider the evaluation of $\mu_{pq}$ by representing $H$
by EGOE(k) and the transition operator ${\cal O}$ by EGOE(t). It
is also assumed that $H$ and $\co$ ensembles are independent.
Now the correlations in  $\mu_{pq}$ arise due to the
non-commutability of $H$ and ${\cal O}$ operators. Firsly it is
seen that all $\mu_{pq}$ with $p+q$ odd will vanish on ensemble
average and also $\mu_{pq}=\mu_{qp}$.  Moreover
$\sigma^2_1=\sigma^2_2=\lan {\cal O}^\dagger {\cal O} H^2 \ran/
\lan {\cal O}^{\dag} {\cal O} \ran={\scriptsize
\l(\barr{c} m \\ k \earr \r) }$. The first non-trivial moment $\mu_{11}$ is,
\be
\barr{rcl}
\zeta=\mu_{11} & = & \lan O^\dagger(t)H(k)O(t)H(k) \ran^m /\l\{
\lan O^\dagger(t)O(t)\ran^m \lan H(k)H(k) \ran^m \r\}
\rightarrow ABAB/(AA)(BB) \nn8
& = & \l(\barr{c} m-t \\ k \earr \r) \l(\barr{c} 
m \\ k \earr \r)^{-1} = 1 - \dis\frac{kt}{m} + \dis\frac{k(k-1)t(t-1)}{2m^2}
\;\;+ O(1/m^3)
\earr
\ee
Let us now consider the cases with $p+q=4$.  The diagrams for
these follow by putting $\cod$ and $\co$ at appropriate places
in the $\lan H^4 \ran$ diagrams,

{\footnotesize
\be
\barr{l}
\mu_{04} =  \l[\lan \cod A A B B \co \ran \;\oplus\; \lan \cod A B B A 
\co \ran \;\oplus\;
\lan \cod A B A B \co \ran \r] / \l[\lan \cod \co \ran (\lan H^2 
\ran)^2 \r] \\
=  2 + \l(\barr{c} m-k \\ k \earr \r) \l(\barr{c} m \\ k \earr \r)^{-1} \nn8
\mu_{13} =  \l[ \lan \cod A A B \co B \ran \;\oplus\; \lan \cod A B B 
\co A \ran  \;\oplus\; \lan \cod A B A \co B \ran \r] /  
\l[\lan \cod \co \ran (\lan H^2 \ran)^2 \r] \\
= \l[ 2 \l(\barr{c} m-t \\ k \earr \r)
\l(\barr{c} m \\ k \earr \r) + \l(\barr{c} m-k \\ k \earr \r) 
\l(\barr{c} m-t \\ k \earr \r) \r] \l(\barr{c} m \\ k \earr \r)^{-2} \nn8
\mu_{22} =  \l[ \lan \cod A A \co B B \ran \;\oplus\; \lan \cod A B 
\co B A \ran \;\oplus\; \lan \cod A B \co A B \ran \r] / 
\l[\lan \cod \co \ran (\lan H^2 \ran)^2 \r] \\
= \l[ \l(\barr{c} m \\ k \earr \r)^2 + 
\l(\barr{c} m-t \\ k \earr \r)^2 + \l(\barr{c} m-k-t \\ k \earr \r)
\l(\barr{c} m-t \\ k \earr \r) \r] \l(\barr{c} m \\ k \earr
\r)^{-2} \nn8
k_{04}=k_{40}=\mu_{40} -3 =  \l(\barr{c} m-k \\ k \earr \r) \l(\barr{c} m
\\ k \earr \r)^{-1} - 1 =  -\dis\frac{k^2}{m} + \dis\frac{k^2(k-1)^2}{2m^2}\;\; +
O(1/m^3) \nn8
k_{13}=k_{31}=\mu_{31}-3\mu_{11} =  \zeta k_{04} \nn8
k_{22}=\mu_{22}-2\zeta^2 -1 = \zeta^2\l\{\l(\barr{c} m-k-t
\\ k \earr \r) \l(\barr{c} m-t \\ k \earr \r)^{-1} - 1 \r\} \\
=  -\dis\frac{k^2}{m} + \dis\frac{k^2\l[(k-1)^2+4kt-2t \r]}{2m^2}\;\; +
O(1/m^3)
\earr
\ee
}

Further results and numerical tests etc. are given in
\cite{Fr-88}.  Finally it is worth mentioning that the EGOE
smoothed Gaussian forms for the state densities (6) and the
bivariate strength densities (8) together with the result that local
strength fluctuations for EGOE (just as for GOE) are of
Porter-Thomas type, are used recently to derive the EGOE
formulas for information entropy ($S^{info}$) and number of
principal components (NPC), which are measures of complexity and
chaos, in transition strength distributions \cite{Ks-98}.

\begin{center}
{\bf 4. EE(1+2) and quantum chaos: transition strength sums}
\end{center}

The principal reason for EGOE receiving considerable attention
in the last 3-4 years is due to the fact that: {\it there is a newly emerging
understanding that in the chaotic domain of isolated finite
interacting many particle systems smoothed densities (they
include strength functions) define the statistical description
of these systems and these densities follow from embedded random
matrix ensembles}. This conjucture is based on the calculations {\bf
for transition strength sums} 
\begin{center}
\begin{tabular}{ll}
\epsfxsize 3in
\epsfysize 4in
\epsfbox{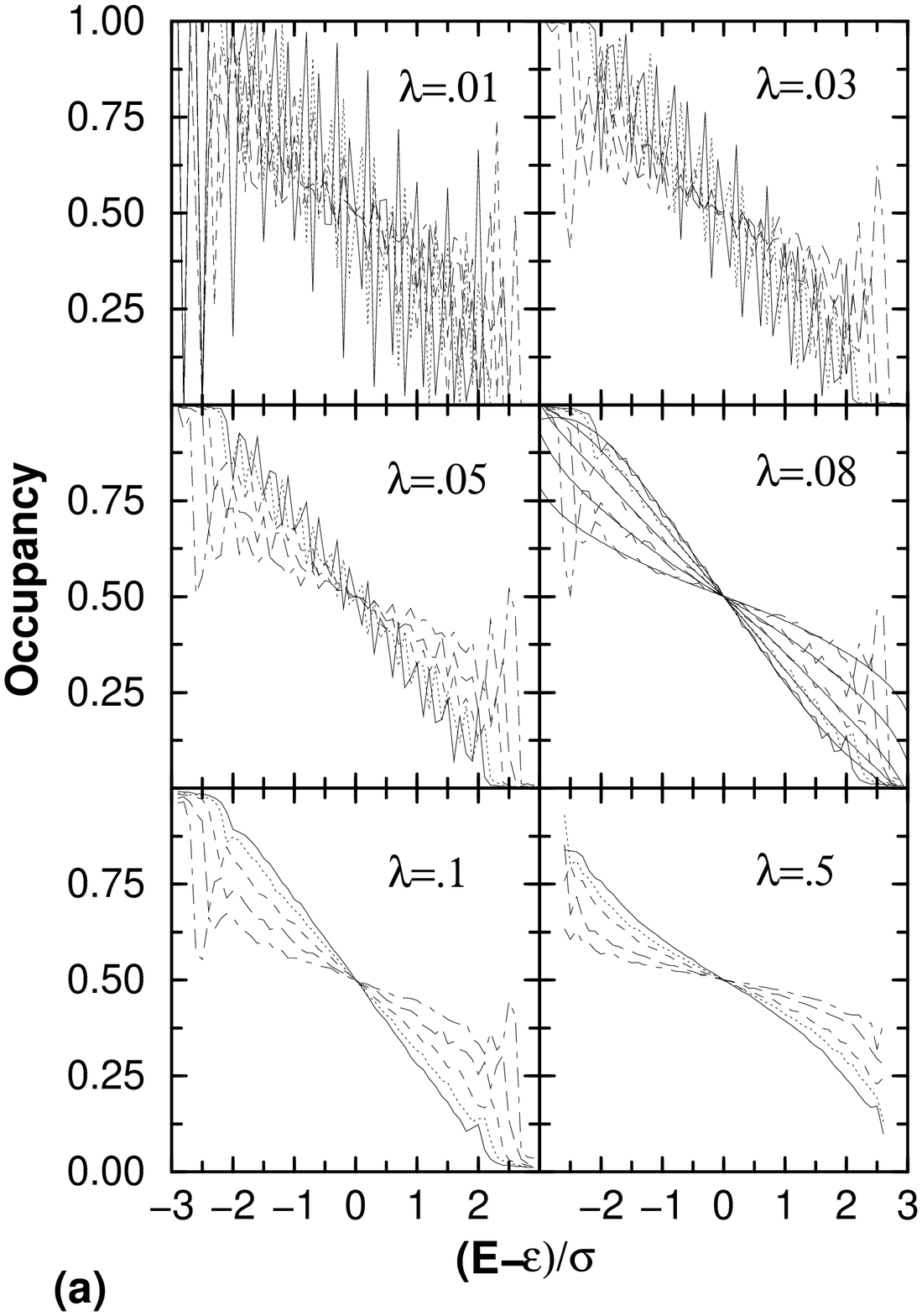}
&
\epsfxsize 3in
\epsfysize 4in
\epsfbox{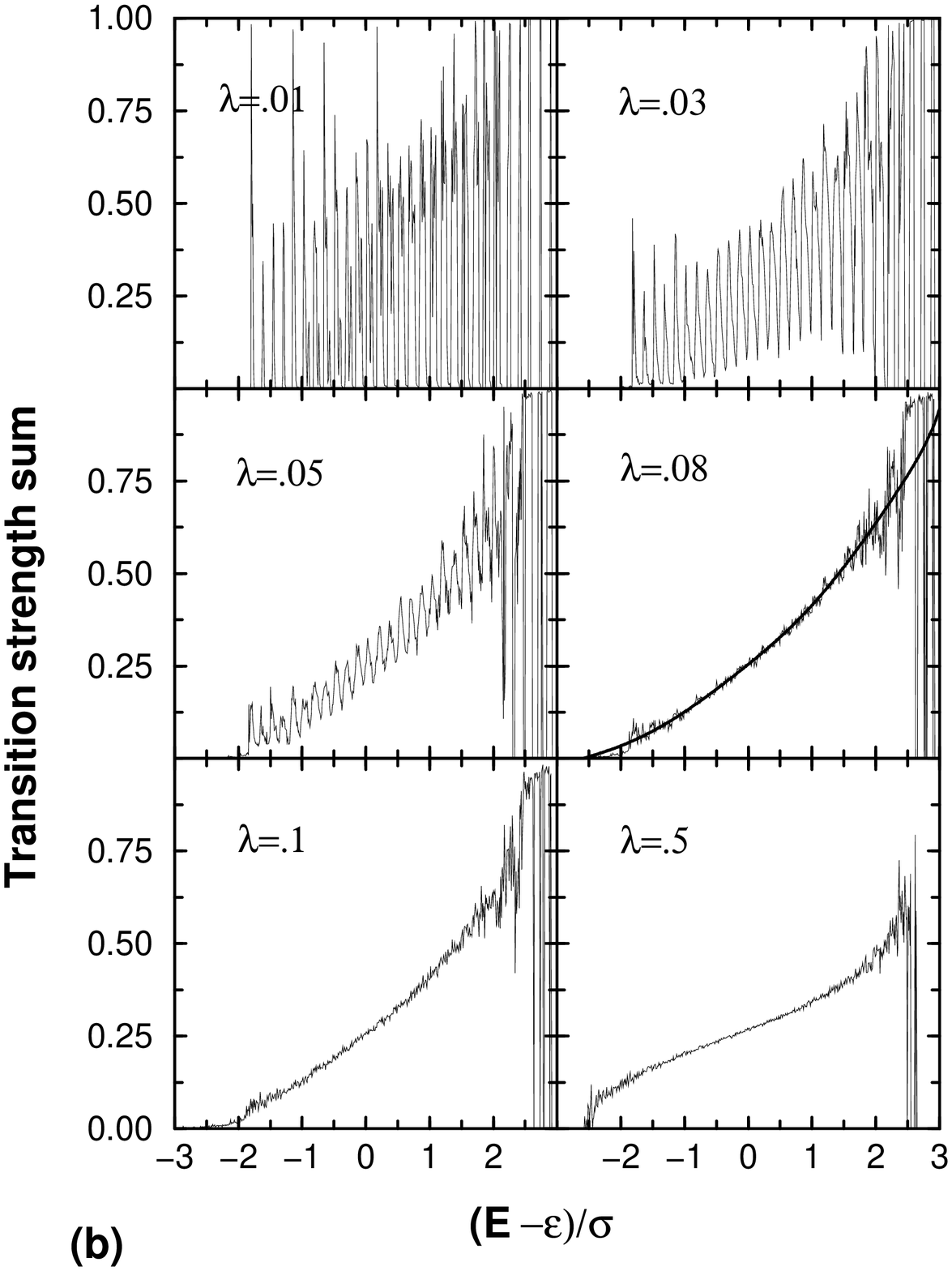} 
\end{tabular}
\end{center}
\vskip -1cm
\noindent {\bf Fig. 1.} {\small {\bf (a)} Occupation numbers for
a 25 member EE(1+2) ensemble, defined by the hamiltonian
$h(1)+\lambda \{V(2)\}$, in the 924 dimensional $N=12$, $m=6$
space. Results are shown for the lowest 5 single particle states
and for six values of $\lambda$. In the calculations occupation
numbers are averaged over a bin size of 0.1 in
$(E-\epsilon)/\sigma$; $\epsilon$ is centroid and $\sigma$ is
width. The spectra of all the ensemble members are first zero
centered and scaled to unit width and then the ensemble average
is carried out. The estimate of \cite{Ja-97} gives
$\lambda_c \sim  0.05$ for order-chaos border in the present
EE(1+2) example. It is clearly seen that once chaos sets in, the
occupation numbers take stable smoothed forms. For
$\lambda=.08$, the smooth forms are well represented by ratio of
Gaussians (smooth curves in the figure).  {\bf (b)} same as (a)
but for the transition strength sum generated by the one-body
transition operator $a^\dagger_2 a_9$. The bin size used here is
0.01.}

\vskip 0.5cm 
\noindent carried out using different types of interpolating
hamiltonians generating order-chaos transitions in many different systems: (i)
for occupancies using a 20 member EE(1+2) in 330 dimensional $N=11,m=4$ space
with $h(1)$ defined by the single particle enegies $\epsilon_i = i+(1/i)\;;\;i
=  1,2,\ldots,11$ and $V(2)$ is EGOE(2) \cite{Fl-97}; (ii) for occupancies
using the four interacting electrons Ce atom \cite{Fl-99}; (iii) for
occupancies using a symmetrized coupled two-rotor model \cite{Ca-98}; (iv)
using two different types of interpolating ensembles in nuclear shell model for
$^{24}$Mg for the Gamow-Teller strength sums and for occupancies \cite{Ko-99}. 
Most significant conclusion of all these studies is that transition strength
sums (note that occupancies are one-particle transfer strength sums) show quite
different behaviour in regular and chaotic domains of the spectrum.  In order
to make this argument clear, calculations are carried out using EE(1+2) as in
(i) but for $N=12$ and $m=6$. The EGOE gives via (8) that the transition
strength sum density $I_{\cod \co}(E)=\lan\lan {\cal O}^\dagger {\cal O}
\delta(H-E) \ran\ran$, which is a marginal density of the bivariate strength
density, is a Gaussian. Therefore, using (6), it is immediately seen that
transition strength sums vary with excitation energy as ratio of Gaussians. 
Fig. 1a shows results for occupancies and Fig. 1b for transition strength sums
for the one-body operator $a^\dagger_2 a_9$ calculated for various values of
the interpolating parameter $\lambda$ in the EE(1+2) hamiltonian.  From Figs.
1a,b it is clearly seen that below the region of onset of chaos transition
strength sums show strong fluctuations (in the regular ground state domain
perturbation theory applies). In this region there is no equilibrium
distribution for $I_{\cod \co}(E)$, $I(E)$ and other densities.  However in the
chaotic domain (there are methods for determining the critical $\lambda_c$ that
marks order-chaos border \cite{Ko-99,Ja-97}) the densities can be replaced by
their smoothed forms.  Note that fluctuations in strength sums are basically
given by 1/NPC. Therefore there is a statistical mechanics, defined by various
smoothed densities and they are given by EGOE, operating in the quantum chaotic
domain of isolated finite interacting particle systems (this is also the
essence of statistical nuclear spectroscopy \cite{Fr-88,Kk-89,Km-96} and also
statistical spectroscopy in atoms \cite{Fl-99}); see also \cite{Ko-99}. In fact
in favourable situations, it is possible to introduce thermodynamic concepts
(effective temperatures and chemical potentials etc.) in the chaotic domain
\cite{Fl-99}. 

\begin{center}
{\bf 5. EE(1+2) results for matrix elements of one-body
transition operators: role of the mean-field basis and strength
functions}
\end{center}

{\bf 5.1 Strength functions: transition from Breit - Wigner to
Gaussian form}

Decomposing the $m$-particle spaces into subspaces $\Gamma$ (say
defined by the irreps of a group structure in the Hilbert space
defined by $(N,m)$), i.e. $m \rightarrow \sum\;\Gamma$, gives,
\be  
\barr{c}  
I(E) = \lan\lan\delta(H-E)\ran\ran = \dis\sum_\Gamma \lan\lan
\delta(H-E)\ran\ran^\Gamma = \dis\sum_\Gamma I^\Gamma(E)\;;\;\;I^{\Gamma}(E)
\longrightarrow {\overline{I^{\Gamma}(E)}} = I^{\Gamma}_{\cal G}(E) \nn8 
I_K(E) = \lan\lan K\delta(H-E)\ran\ran = \dis\sum_\Gamma \lan\lan
K\delta(H-E)\ran\ran^\Gamma = \dis\sum_\Gamma I_K^\Gamma(E)\;;\;\;
I^\Gamma_K(E) \longrightarrow {\overline{I^\Gamma_K(E)}} =  I^\Gamma_{K:{\cal
G}}(E)  
\earr  
\ee  
There are plausible arguments in favour of the Gaussian forms
in (11) for EGOE. The decompositions in (11) are related to the well known 
`strength functions' (also called local spectral density of states
(LDOS) in literature) which are basic ingredients of a many particle
system. Given a compound state $\phi_k$, the probability of its decay into
stationary states $\psi_E$ (generated by $H$) is given by 
$\l|\lan \phi_k \mid \psi_E\ran\r|^2$. Then the strength function $F_k(E)$ is,
\be
\l|\l.\phi_k\ran\r. = \dis\sum_E\;C^E_k\;\l|\l.\psi_E\ran
\r.\;\;;\;\;\;\;
F_k(E) = \dis\sum_{E^\pr}\; \l|C^{E^\pr}_k\r|^2\;\delta(E-E^\pr) = \lan
\delta(H-E) \ran^k
\ee
If we consider a set of states $\phi_k$ that belong to a irrep
$\Gamma$ as in (11), then the average strength function
$F_{\Gamma}(E)$ is nothing but the partial densities
$I_\Gamma(E)$ in (11); $I_\Gamma(E) = \sum_{k \varepsilon
\Gamma}\,F_k(E)$ and $F_{\Gamma}(E)= I_{\Gamma}(E)/d(\Gamma)$.
A quite different and useful way to look at strength functions
is to think of $\phi_k$ as a compound state generated by the
action of a transition operator ${\cal O}$ on a state
$\psi_{E_k}$ (for example ground state),
\be
\barr{c}
\l|\l.\phi_{E_k}\ran\r. = \dis\frac{{\cal O}\;\l|\l.\psi_{E_k}\ran\r.}{
\l[\lan\psi_{E_k} \mid {\cal O}^\dagger {\cal O} \mid \psi_{E_k}
\ran\r]^{1/2}}\;\;;\;\;\;\lan \phi_{E_k}\mid\phi_{E_k}\ran = 1 \nn8
\l|\l.\phi_{E_k}\ran\r. =
\dis\sum_{E}\;C^E_{E_k}\;\l|\l.\psi_E\ran\r. \;\;\;\Rightarrow \;\;\;
F_{k;{\cal O}}(E) = \dis\frac{\lan\lan {\cal O}^\dagger \delta(H-E)
{\cal O} \delta(H-E_k) \ran\ran}{\lan\lan {\cal O}^\dagger {\cal
O} \delta(H-E_k) \ran\ran} 
\earr
\ee
\newpage

\begin{center}
\epsfxsize 3.5in
\epsfysize 4in
\epsfbox{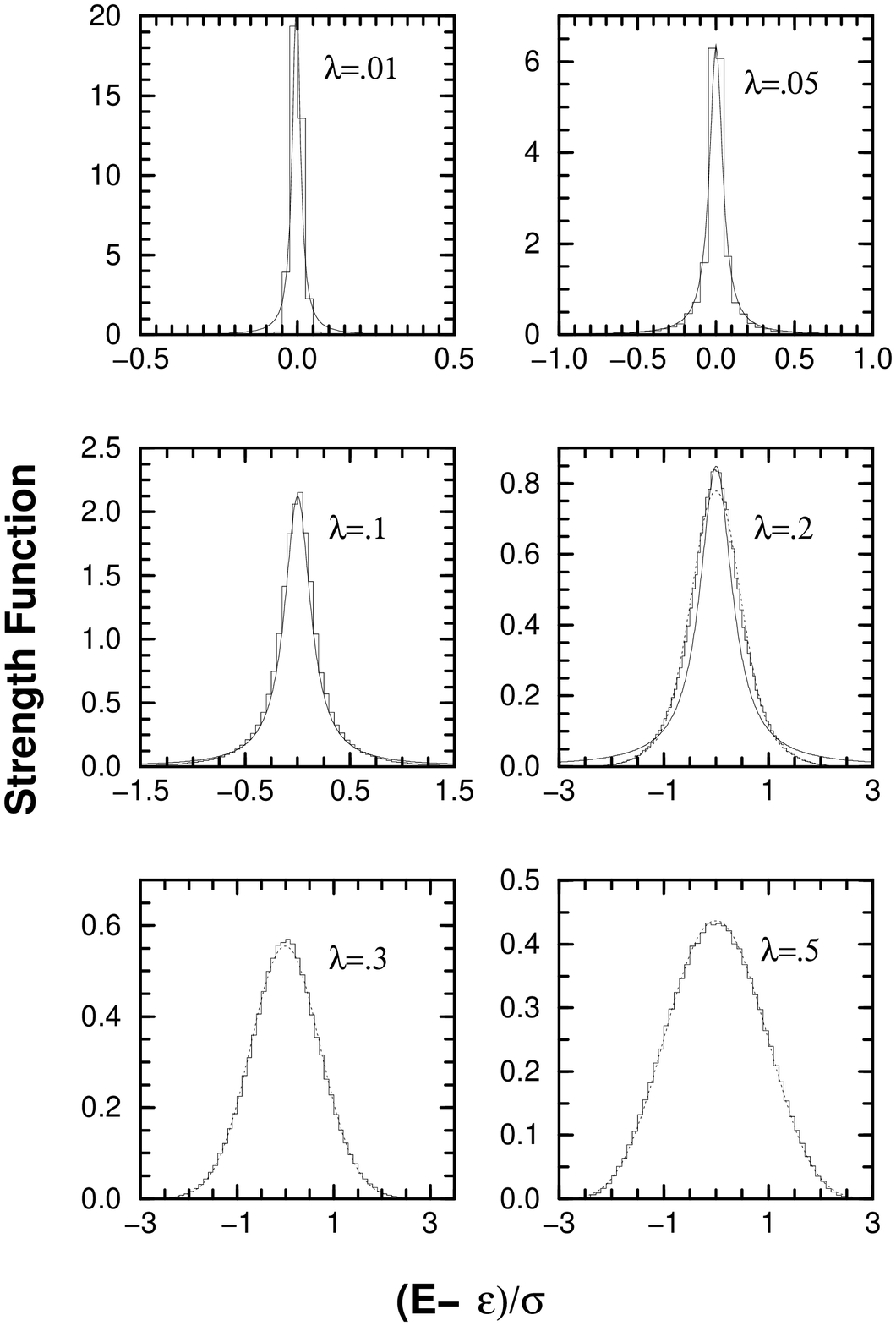}
\end{center}
\vskip -1.5cm
\noindent {\bf Fig. 2.} {\small 
Strength functions ($F_{\ce_k}(E)$) for
the EGOE(1+2) ensemble defined in Fig. 1. The $\l.\l| k
\r. \ran$ states are the mean-field $h(1)$ states defined by the
distribution of $m$ particles in the $N$ single particle states
and their energies $\ce_k$ are $\ce_k=\lan k | h(1) + \lambda \,
V(2) | k \ran$. In the calculations $E$ and $\ce_k$ are zero centered for
each member and scaled by the spectrum ($E$'s) width $\sigma$;
${\hat{E}}_k = (\ce_k-\epsilon)/\sigma$ and ${\hat{E}} =
(E-\epsilon)/\sigma$.  For each member $|C_k^E |^2$ are summed
over the basis states $\l.\l| k \r. \ran$ in the energy window
${\hat{E}}_k \pm \Delta$ and then ensemble averaged
$F_{{\hat{E}}_k} ({\hat{E}})$ vs ${\hat{E}}$ is constructed as a
histogram; the value of $\Delta$ is chosen to be 0.025 for
$\lambda < 0.1$ and beyond this $\Delta=0.1$.  Results are shown
for $\lambda=$ 0.01, 0.05, 0.1, 0.2, 0.3 and 0.5. The histograms are
EGOE(1+2) results, continuous curves are BW fit and dotted
curves are Edgeworth corrected Gaussians (ED); the ED
incorporates \cite{Ke-69} skewness ($\gamma_1$) and excess
($\gamma_2$) corrections. For $\lambda$ = 0.01, 0.05 and 0.1
only BW, for 0.2 both BW and ED and for 0.3 and 0.5 only ED are
shown. In the figure all the results are for ${\hat{E}}_k =0$.}

\vskip 0.5cm
Starting with the EE(1+2), it is to be expected that for
sufficiently large values of $\lambda$ in (5), EGOE($k$) description
should be valid and therefore applying (8) gives the shape of
the strength function to be Gaussian (conditional density of a
bivariate Gaussian is a Gaussian) and hence its width $\sigma_k$
is independent of the energy $E_k$ or $k$. These results are indeed seen
in numerical calculations. The important question is how
$F_k(E)$ changes as $\lambda$ is varied. The standard form,
normally employed in many applications, for strength functions
is the Breit-Wigner (BW) form characterized by a spreading width
$\Gamma_k$,
\be
F_{k:BW}(E) =
\dis\frac{1}{2\pi}\;\dis\frac{\Gamma_k}{(E-{\overline{E_k}})^2 +
\Gamma^2_k/4}
\ee
where ${\overline{E_k}} = \lan k \mid H \mid k \ran = \int^\infty_{-\infty}
F_k(E) E \;dE$. It is expected that BW form and the Gaussian
form should appear as $\lambda$ is varied and this question is
investigated in: (i) using EE(1+2) and constructing $F_k(E)$ for
various values of $\lambda$, here $\phi_k$ are the mean-field
($h$) basis states (only final result of this study is reported in 
\cite{Fl-97}); (ii) by carrying out nuclear
shell model calculations with $H_\lambda=h+\lambda\,V$ for
$^{28}$Si with $\phi_k$ chosen to be mean-field basis states
\cite{Bz-96}; (iii) using the three-orbital Lipkin-Meshkov-Glick
model \cite{Wa-98}; (iv) using a symmetrized coupled two-rotor
model \cite{Ca-98}.  Results from these studies show that when
the system is chaotic (so that level and strength fluctuations
follow from GOE), strength functions take Gaussian form and
order-chaos transition implies for $F_k(E)$ transition from BW
to Gaussian,
\be
order \rightarrow chaos \;\;\;\Rightarrow \;\;\; F_{k:BW}(E) 
\longrightarrow F_{k:{\cal G}}(E)
\ee
For further understanding of the nature of $F_k(E)$ we performed
EGOE(1+2) calculations in the 924 dimensional $N=12$ and $m=6$
space and the results are shown in Fig. 2. From the figure it is
clearly seen that there is BW to Gaussian transition and the
value of the interpolating parameter $\lambda= \lambda_{F_k}$
for onset of this transition is $\lambda_{F_k}
\approx 0.2$.  This should be compared with $\lambda = \lambda_c
\approx 0.08$ for onset of chaos in level fluctuations, occupancies
and strength sums of one-body operators in the present EGOE(1+2)
example (Fig. 1).  A similar BW to Gaussian
transition is seen in \cite{Bz-96,Wa-98} and in the nuclear
shell model example considered in \cite{Bz-96} $\lambda_c \sim
0.3$ and $\lambda_{F_k} \sim 0.6$. Thus the BW form for $F_k(E)$
extends into the chaotic domain ($\lambda > \lambda_c$) and the
transition to Gaussian shape takes place in the second chaotic layer 
defined by $\lambda > \lambda_{F_k}$. Full understanding of $\lambda_{F_k}$
is at present lacking.

{\bf 5.2 matrix elements of one-body transition operators in
the chaotic domain}

Let us consider a one-body transition operator written in occupation number
representation $\co = \epsilon_{\alpha \beta}a^\dagger_\alpha a_\beta$ and its
matrix elements $\l| \lan E_f \mid \co \mid E_i \ran \r|^2$ in many-particle
eigenstates expanded in the mean-field basis states $\l| k_i \ran$ are (with
$\epsilon_\alpha$ being single particle energies and $\ce_i$ being the
mean-field basis states energies), 
\be 
\barr{c} \l| \lan E_f \mid \co \mid E_i
\ran \r|^2 = \l\{\dis\sum_{k_i\;k_f} C^{E_i}_{k_i}\;C^{E_f}_{k_f}\;\lan k_f
\mid \co \mid k_i \ran \r\}^2 = diag + offdiag \nn8 = \dis\sum_{k_i\;k_f} \l|
C^{E_i}_{k_i} \r|^2\;\l|C^{E_f}_{k_f}\r|^2\;\l|\lan k_f \mid \co  \mid k_i \ran
\r|^2  + \dis\sum_{k_i \neq k_i^\pr\;,\;k_f \neq k_f^\pr}
C^{E_i}_{k_i}\;C^{E_i}_{k^\pr_i}\;C^{E_f}_{k_f}\;C^{E_f}_{k^\pr_f}\; \lan k_f
\mid \co \mid k_i \ran \lan k^\pr_f \mid \co \mid k^\pr_i \ran \nn8 \earr 
\ee
The $diag$ term in (16) involves only the strength functions (see (12)) and
also for this term, for a given $k_f$ and $k_i$ only one  $\epsilon_{\alpha
\beta}$ in $\co$ will contribute. Following \cite{Km-96} it is easily seen
that, 
\be \l| \lan E_f \mid \co \mid E_i \ran \r|^2_{diag} =
\dis\sum_{\alpha\;\beta} \l|\epsilon_{\alpha \beta}\r|^2\; \l\{\dis\sum_{\ce_i}
\lan n_\beta (1-n_\alpha) \ran^{\ce_i} \l|C^{E_i}_{ \ce_i}\r|^2 \;\l|
C^{E_f}_{\ce_f=\ce_i-\epsilon_\beta+\epsilon_\alpha}\r|^2 \r\} 
\ee 
Assuming
that $\lan n_\beta (1-n_\alpha) \ran^{\ce_i}$ do not vary much over the number
of principal components (say $N_{eff}$ is NPC for the mean-field basis states),
it can be replaced by its corresponding average (this is verified using EE(1+2)
in \cite{Fl-96}). Then, 
\be 
\l| \lan E_f \mid \co \mid E_i \ran \r|^2_{diag} =
\dis\sum_{\alpha\;\beta} \l|\epsilon_{\alpha \beta}\r|^2\;\lan n_\beta
(1-n_\alpha)  \ran^{E_i}\; \l\{\dis\sum_{\ce_i}  \l|C^{E_i}_{\ce_i}\r|^2 \;\l|
C^{E_f}_{\ce_f=\ce_i-\epsilon_\beta+\epsilon_\alpha}\r|^2 \r\} 
\ee 
The
$|C|^2$'s in (18) are nothing but strength functions. With the assumption that
strength functions are a function of $(E-\ce_i)/s_i$ where $s_i$ is a scale
parameter (spreading width $\Gamma_i$ for BW and the spectral width $\sigma_i$
for Gaussian), we have the results (with $\overline{D(E)}$ denoting mean
spacing), 
\be 
\barr{c} F_k(E) = (s_k)^{-1} f((E-\ce_k)/s_k)\;,\;\;\;\int F_k(E)
dE =1 \nn8 \l|C^E_k\r|^2 = {\overline{D(E)}}\; F_k(E)\;\Rightarrow \l(N_{eff}
f(0)\r)^{-1} f((E-\ce_k)/s_k)\;,\;\;\; {\overline{D(E)}}=\l[N_{eff}
f(0)\r]^{-1}\;\overline{s} \nn8 \dis\sum_k \l|C^E_k\r|^2 \longrightarrow
{\large{\int}} \l|C^E_k\r|^2 \dis\frac{d \ce_k}{{\overline{D(E)}}} = \int
F_k(E) d\ce_k \earr 
\ee 
Substituting $F$'s for $|C|^2$ in (18) and replacing
the $\ce_i$ summation by an integral, both using (19), give the final result in
terms of occupancies and the mean spacings, 
\be 
\barr{c} \l| \lan E_f \mid \co
\mid E_i \ran \r|^2_{diag} = \dis\sum_{\alpha, \beta} \l|\epsilon_{\alpha
\beta}\r|^2\;\lan n_\beta (1-n_\alpha)  \ran^{E_i} {\overline{D(E_f)}}\;\int
F_{\ce_i}(E_i)\; F_{ \ce_f=\ce_i-\epsilon_\beta+\epsilon_\alpha}(E_f)\;\;d\ce_i
\nn8 = \dis\sum_{\alpha, \beta} \l|\epsilon_{\alpha \beta}\r|^2\;\lan n_\beta
(1-n_\alpha)  \ran^{E_i} {\overline{D(E_f)}}\;\cf(\Delta,s_i,s_f)\;;\;\;\;\;\;
\Delta=E_f-E_i+\epsilon_\beta-\epsilon_\alpha \nn8
\cf(\Delta,\Gamma_i,\Gamma_f)_{BW}= \dis\frac{1}{2 \pi} 
\dis\frac{\Gamma_i+\Gamma_f}{\Delta^2+ (\Gamma_i+\Gamma_f)^2/4}\nn8
\cf(\Delta,\sigma_i,\sigma_f)_{Gauss}= \dis\frac{1}{\dis\sqrt{ 2 \pi
(\sigma_i^2 + \sigma_f^2)}}\;exp-\dis\frac{ \Delta^2}{2(\sigma_i^2 +
\sigma_f^2)} \earr 
\ee 
Eq. (20) with $\cf_{BW}$ was derived earlier
\cite{Fl-99,Fl-96}. The procedure used in deriving $diag$ term is equivalent to
starting with $H=h(1)+V(2)$, first generating the non-interacting paricle (NIP)
bivariate strength density $I^h_{biv; \co}(x,x^\prime)$ due to $h(1)$  (using
the results of \cite{Km-96}), then convoluting the NIP spikes with independent
spreading functions, due to $V(2)$, at the two energies and finally simplifying
as in (20). This assumes that the spreadings are uncorrelated. However, the
$offdiag$ term in (16) arises due to correlations and the bivariate correlation
coefficient $\zeta=\lan \cod V \co V \ran/\lan \cod \co\ran \lan V V \ran$,
takes them into account, i.e. the spreading function $\rho^V_{biv;
\co}(y,y^\prime)$ in  
$$ 
I^H_{biv;\co}(E,E^\prime) = I^h_{biv; \co} \otimes
\rho^V_{biv; \co} [E,E^\prime] 
$$ 
is not $\rho^V_{{\cal G};i}(y) \rho^V_{{\cal
G};f}(y^\prime)$ but  it is
$\rho^V_{biv;\co}(y,y^\prime;0,0,\sigma_i,\sigma_f,\zeta)$.   Following
\cite{Km-96} it is easily seen that even with $\zeta$ the final result is same
as in (20) but with a modified $\cf$, 
\be
\cf(\Delta,\sigma_i,\sigma_f,\zeta)_{biv-{\cal G}}= \dis\frac{1}{\dis\sqrt{ 2
\pi (\sigma_i^2 + \sigma_f^2 - 2 \zeta \sigma_i \sigma_f )}}\;exp-\dis\frac{
\Delta^2}{2(\sigma_i^2 + \sigma_f^2 - 2 \zeta \sigma_i \sigma_f)}  
\ee
Therefore for $\Delta=0$ and $\zeta \rightarrow 1$, there is enhancement in the
matrix elements compared to the $diag$ approximation.  This is indeed seen in
EE(1+2) calculations earlier \cite{Fl-96} and now it is clear that with $\zeta$
in $\rho^V_{biv;\co}$ we have the proper theory for transition matrix elements
of one-body operators (the theory with (21) is expected to
operate in the chaotic domain defined by $\lambda >
\lambda_{F_k}$ and $\lambda_{F_k}$ is defined in Section 5.1) . 
In fact for $\sigma_i \sim \sigma_f$, the enhancement is
$1/\sqrt{(1-\zeta)}$. Note that for $\zeta=0$ we get back the $diag$
approximation which is a GOE result. Also as noted in \cite{Fl-96}, the
enhancement grows with $m$ as $\zeta$ grows with $m$ (see (9)). Applications of
the results in (20,21) and their variants \cite{Km-96}, for nuclei and atoms
are given in \cite{Km-96,Mk-98} and \cite{Fl-99,Gf-99} respectively.

\begin{center}
{\bf 6. Conclusions}
\end{center}

In this article an attempt is made to give an overview of the subject of
embedded random matrix ensembles (classical EGOE and its various deformations)
for complexity and chaos in interacting particle systems.  It should be clear
that generic embedded ensembles results are relevant in the quantum chaotic
domain of isolated finite  interacting many particle systems such as nuclei,
atoms, molecules, atomic clusters, quantum dots etc. and therefore large scale
explorations  of deformed embedded ensembles are called for.

\ed